\documentclass[%
  10pt,amsmath,amssymb,floatfix,
  aps,prb,twocolumn
]{revtex4-2}

\usepackage{ragged2e}
\usepackage{amsmath}
\usepackage{graphicx}
\usepackage{dcolumn}
\usepackage{bm}
\usepackage[utf8]{inputenc}
\usepackage{subcaption}
\usepackage{hyperref}

\renewcommand{\bibsection}{\section*{References}}

\begin{document}


\title{\Large Tunnelling probe-based characterisation of the sp${}^3$ dangling bond \\
on the H--C(100):$2\times1$ surface}
\vspace{1em}

\author{Lachlan Oberg$^{1,2}$}\email{Corresponding author: l.oberg@qut.edu.au}
\author{Yi-Ying Sung$^3$}
\author{Cedric Weber$^4$}
\author{Marcus W. Doherty$^4$}
\author{Christopher I. Pakes$^3$}

\affiliation{%
  $^1$ Department of Quantum Science and Technology, Research School of Physics,
  Australian National University, Canberra 2601, Australia.%
}
\affiliation{%
  $^2$ School of Civil \& Environmental Engineering, Faculty of Engineering,
  Queensland University of Technology, Gardens Point Campus, Brisbane, QLD 4000, Australia.%
}
\affiliation{%
  $^3$ Department of Mathematical and Physical Sciences, La Trobe University, Victoria 3086, Australia.%
}
\affiliation{%
  $^4$ Quantum Brilliance Pty Ltd, 60 Mills Road, Acton, ACT 2601, Australia.%
}

\date{September 30, 2026}

\begin{abstract}

The sp${}^3$ dangling bond on the diamond surface plays a critical role in the performance and fabrication of diamond quantum technologies. For the former, the magnetic and electric properties of this defect can impede the performance of quantum sensors and computers. For the latter, the chemical properties of the dangling bond are integral to proposed methods for bottom-up fabrication of scalable diamond quantum devices. In pursuit of high-performance, scalable diamond quantum technology, tunnelling probe-based techniques offer the ability to create and modify the sp${}^3$ dangling bond with atomic-scale precision. However, these capabilities cannot be realised either deterministically or at scale without a means of identifying the sp${}^3$ dangling bond amidst the myriad of other defects on the diamond surface. Consequently, in this work we provide a comprehensive experimental and theoretical framework for STS-based characterisation of the sp${}^3$ defect on the H-terminated (100) diamond surface. This capability provides the foundation for future tunnelling probe studies in the modification of dangling bonds.

\end{abstract}

\maketitle

\section{Introduction}

\begin{figure}
    \centering
    \begin{subfigure}[b]{0.45\textwidth}
        \includegraphics[width=\textwidth]{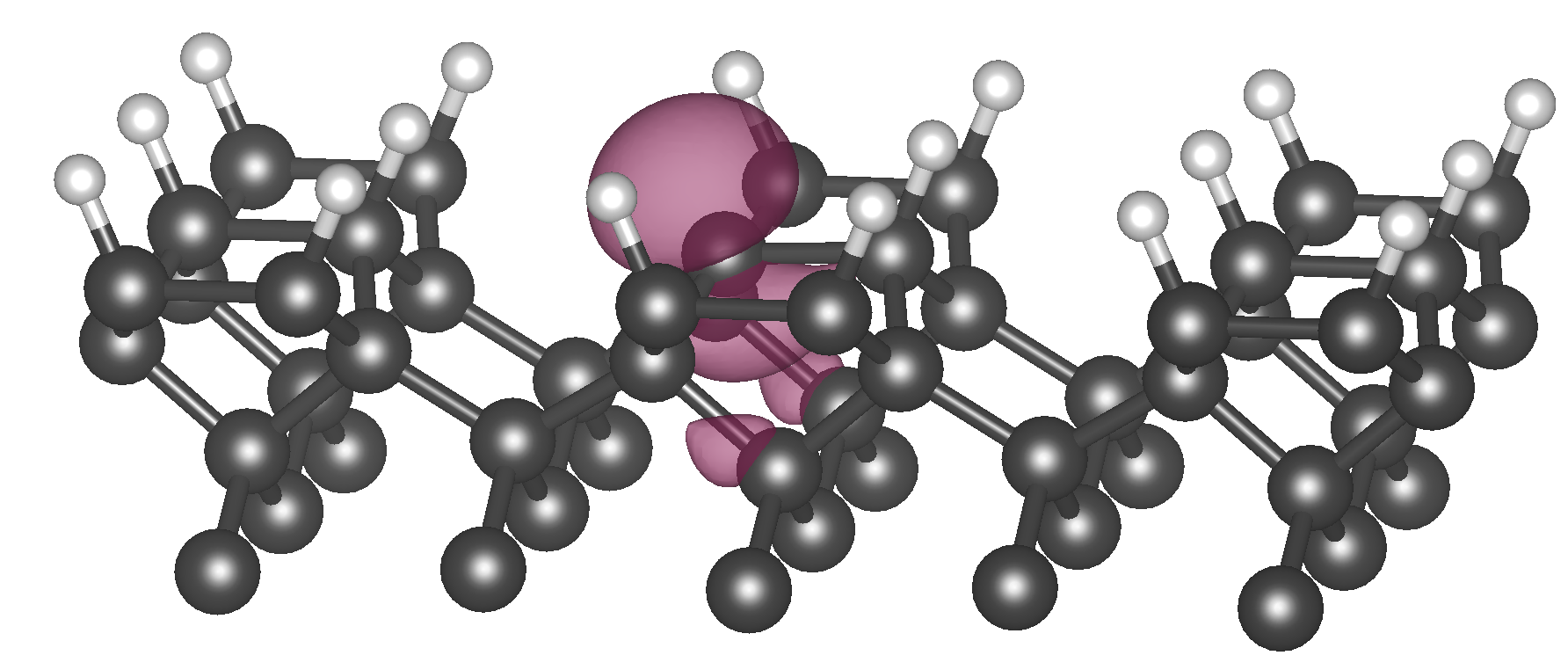}
        \caption{ Side view.}
    \end{subfigure}

    \begin{subfigure}[b]{0.45\textwidth}
        \includegraphics[width=\textwidth]{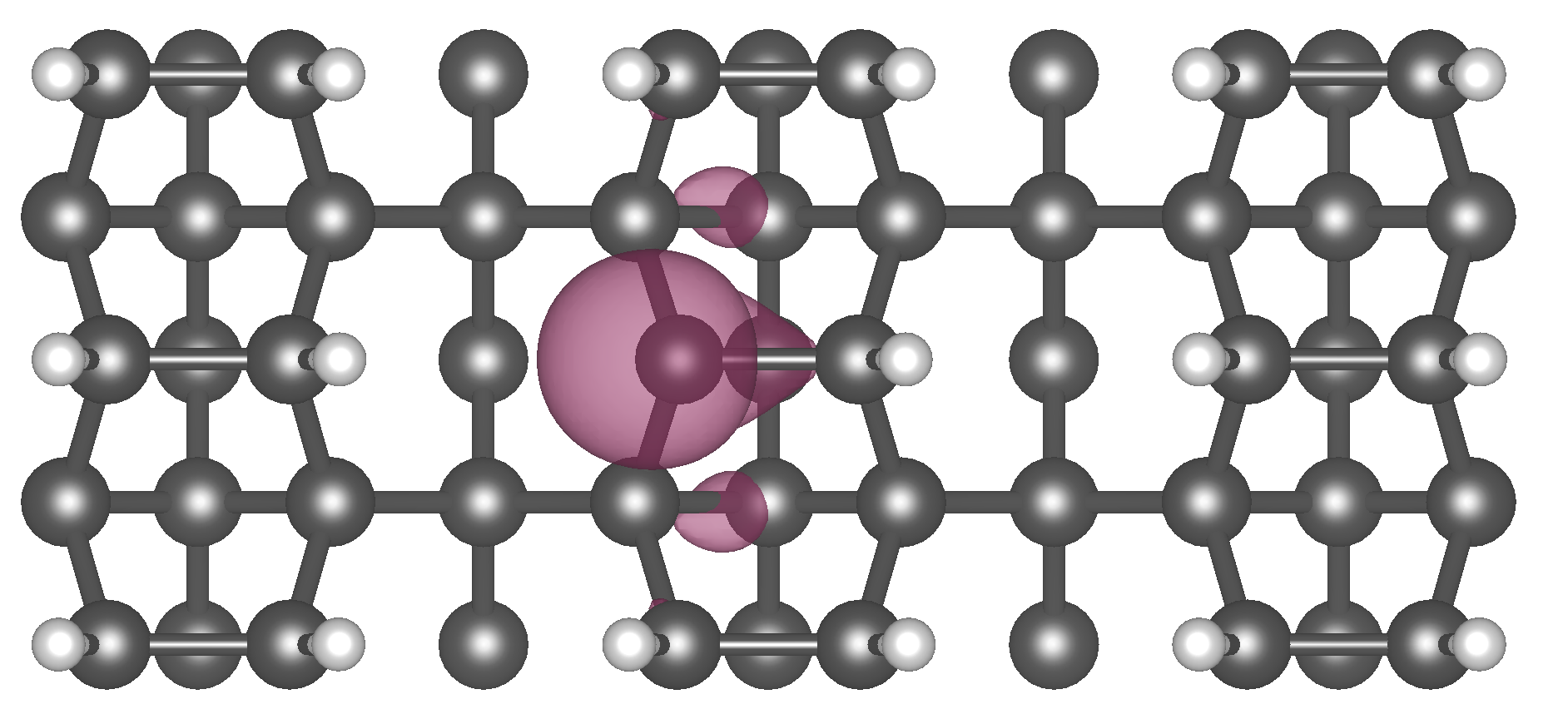}
        \caption{ Top view.}
    \end{subfigure}

    \caption{\justifying Atomic structure of the sp${}^3$ dangling bond on the H--C(100):$2\times1$ surface. Hydrogen ions are white and carbon atoms are grey. A single hydrogen ion has been removed from the dimerised surface, exposing a bare surface carbon with an sp${}^3$ dangling bond. This dangling bond produces localised states within the diamond band gap which can be visualised using STM. Depicted in purple is the charge density corresponding to an unoccupied mid-gap state.}\label{fig:1}
\end{figure}

The sp${}^3$ dangling bond on the hydrogen-terminated diamond surface has immense significance for diamond quantum technologies. As depicted in Figure~\ref{fig:1}, this defect is formed through the desorption of a terminating H atom and is a common byproduct during diamond chemical vapour deposition (CVD) growth and surface preparation\cite{OBERG2021606,Stacey2019,Sung2025}. On the technologically important (100) and (111) surfaces, the defect possesses an unsatisfied electron at the bare carbon (C) surface site with an sp${}^3$ bond hybridisation. Unfortunately, it exhibits paramagnetic and charge-trapping behaviour which can impede the performance of diamond quantum computers and quantum sensors which rely upon near-surface nitrogen-vacancy (NV) centres\cite{Stacey2019,Sung2025,Chou2023,PhysRevApplied.14.014085,doi:10.1021/acs.nanolett.1c00082,Kim2015,Oberg2019Nanowires}. This is chiefly through the production of magnetic and electric noise that reduces NV coherence times and causes electric field screening during electrometry. 

Fortunately, recent results have demonstrated that the deleterious effects of the sp${}^3$ dangling bond can be neutralised through H capping\cite{Sung2025}. Voltage pulses are applied to a hydrogen-functionalised scanning tunnelling microscope (STM) tip to passivate individual defects with H termination. However, this capping procedure is currently unreliable, with poor control, yield, and a spatial resolution on the order of nanometres. Moreover, due to visual similarities with other benign surface defects, identifying sp${}^3$ dangling bonds for H capping can be challenging. While the defects may be differentiated using high-resolution STM imaging, scanning tunnelling spectroscopy (STS) offers an alternative and direct means for identifying sp${}^3$ dangling bonds, but has not been substantively explored in the literature.

Additionally, recent work has proposed that scalable diamond quantum devices could be fabricated through the targeted creation of sp${}^3$ dangling bonds\cite{Doherty2025,Oberg2024,Oberg_2025}. Termed H desorption lithography (HDL), this process involves applying voltage pulses to an STM tip for the targeted removal of H termination\cite{BobrovHDL}. This produces reactive clusters of dangling bonds which subsequently act as active sites for chemical adsorption of a nitrogen-containing gas. Lastly, the surface is overgrown using CVD to produce bulk NV centres. By using HDL to produce many active sites simultaneously, it may be possible to realise a large-scale diamond quantum device containing many coupled NV centres. However, a key requirement for this multi-step process is validating the production of dangling bonds following lithography. This analysis could be greatly enhanced through STS, which provides a distinct signature that reveals the energetic structure of the sp${}^3$ dangling bond.

Consequently, the aim of this work is to characterise the STS spectra of sp${}^3$ dangling bonds through a combined theoretical and experimental approach. While first-principles calculations have previously determined the defect electronic structure\cite{Sung2025}, this structure has yet to be aligned with features of experimental STS (of which there are limited examples in the literature). Furthermore, performing this alignment is greatly complicated due to band bending on diamond surfaces. Diamond STM/STS typically relies on boron doping to achieve relatively low conductivities under UHV conditions. Electric fields are therefore poorly screened at the diamond surface, resulting in significant band bending\cite{Sung2025}. In contrast to STM/STS on metallic surfaces, the applied bias between tip and sample is \textit{not} commensurate to the energetic translation of the tip Fermi level relative to surface states. Hence, peak positions in STS spectra are not necessarily equivalent to the energy of surface electronic states\cite{Feenstra2007,Feenstra2009}. Careful analysis is therefore required to account for modifications in STS spectra due to band bending.

The outline of this work is as follows. In Section~\ref{sec:methods} we present our experimental and theoretical methodology. This entails STS of an sp${}^3$ dangling bond occurring on an as-grown H--C(100):$2\times1$ surface, calculating the defect electronic structure using density functional theory, and modelling the effects of band bending using electrostatic simulations. Section~\ref{sec:results} presents the resulting experimental spectrum, theoretical band structure, and surface band bending. In Section~\ref{sec:disc} we synthesise these results to assign peaks in the experimental spectrum to electronic states of the defect. Moreover, this assignment process enables estimation of the acceptor concentration and absolute tip-surface distance. The former introduces STS as a bulk characterisation tool whereas the latter is a long-standing challenge in STM. Finally, Section~\ref{sec:conc} concludes that STS in coordination with STM imaging can be used to identify sp${}^3$ dangling bonds.

\section{Methodology}
\label{sec:methods}

\subsection{Experimental}

The sample was a boron-doped single crystal diamond surface epilayer grown by chemical vapour deposition on an underlying type IIa single-crystal diamond with a (100) orientation, purchased from Element Six Ltd. Specific details regarding the preparation methodology are available in Ref.~\cite{Sung2025}. To preserve the high quality of the surface for atom-scale STM imaging, ex situ measurements to estimate the B concentration of the sample used in this study were not undertaken. However, we estimate a boron concentration in the range of $10^{16}$~$\text{cm}^{-3}$ to $10^{20}$~$\text{cm}^{-3}$. This estimate is based on electrical conductivity measurements, x-ray photoelectron spectroscopy, and time-of-flight secondary-ion mass spectrometry performed on similarly prepared samples.

STM and STS were carried out using a SPECS Aarhus UHV--STM operating at room temperature and under ultra-high vacuum (UHV) with a base pressure of $10^{-9}$ mbar. CVD-grown C(100) samples of dimensions $3.5 \times 3.5$ mm${}^2$ were outgassed at 823 K for 1 h to remove airborne surface impurities and produce a small number of dangling bonds. STM images were recorded in constant-current mode at $+2.5$~V sample bias (i.e., imaging the empty states) and with the tunnelling current set to 0.3~nA. Single-point $dI/dV$ spectra were obtained by positioning the STM tip on either the dimer row or the dangling bond sites. The spectra were recorded using a lock-in amplifier (modulated at 583 Hz with 50~mV peak-to-peak amplitude) while ramping the sample bias from +4~V to -2.5~V at constant tip-sample distance. This fixed distance corresponded to tunnelling conditions at $I = 0.30$~nA and $V = 4$~V.

\subsection{Density functional theory}

Density functional theory was performed using the \textit{Vienna Ab initio Simulation Package} (VASP)\cite{Kresse1993,Kresse1996a,Kresse1996b}. The H--C(100):$2\times1$ surface and sp${}^3$ dangling bond were represented using a 419-ion slab with a thickness of 8 carbon layers and lateral dimensions of 17.7 \ \AA\  $\times$ 15.2 \ \AA. The slab geometry was optimised using calculations with the Perdew--Burke--Ernzerhof (PBE) functional\cite{Perdew1996} to a force tolerance of $5\times{10}^{-3}$~eV/\AA \ per ion. We used a plane-wave cut-off of 600~eV, a $4\times4\times1$ $\Gamma$-centred \textit{k}-point mesh, and the electronic convergence tolerance was 10${}^{-4}$~eV. Using the optimised slab geometry, we then performed $\Gamma$-point calculations using the HSE06 functional\cite{Heyd2003} and a constant electric field (treated self-consistently) between $\pm0.3$~V/\AA. Details of the slab model can be found in Ref.~\cite{Sung2025}.

\subsection{Electrostatic modelling}

The effects of band bending during diamond STS were simulated through electrostatic modelling with COMSOL Multiphysics. Poisson's equation was solved via finite element analysis for a space-charge density and geometry appropriate for a metallic tip above the surface of boron-doped diamond. Extensive details regarding this modelling are provided in Refs.~\cite{Sung2025,Oberg2024}. Simulations were performed for a range of tip-sample separation distances and boron concentrations. The key results of this modelling are the voltage drop ($V_\text{drop}$, defined as the voltage difference between the tip and sample surface) and electric field at the diamond surface at the position directly beneath the tip apex. These two values define all aspects of band bending needed to interpret the STS spectra, and were calculated as a function of applied bias ($V_\text{bias}$) between the tip and sample backplate.

\section{Results}
\label{sec:results}

\subsection{Experimental STS}

\begin{figure}[tbp]
    \centering
    \begin{subfigure}[b]{0.47\textwidth}
        \includegraphics[width=\textwidth]{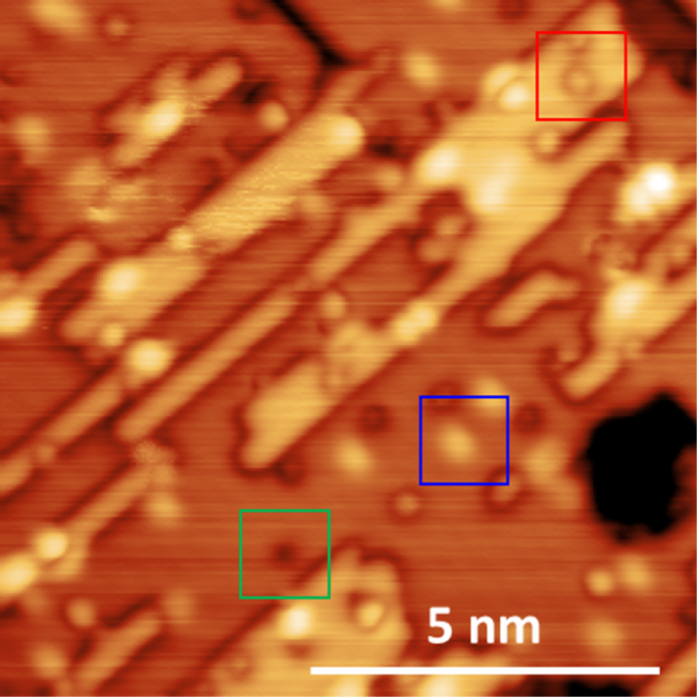}
        \caption{\justifying STM image of the H--C(100):$2\times1$ surface displaying a myriad of defects amongst the characteristic dimer rows. Indicated in green is a single vacancy, in red a CH${}_2$ bridge, and in blue the sp${}^3$ dangling bond. It is characterised by a bright asymmetric feature.}
    \end{subfigure}

    \begin{subfigure}[b]{0.47\textwidth}
    \includegraphics[width=\textwidth]{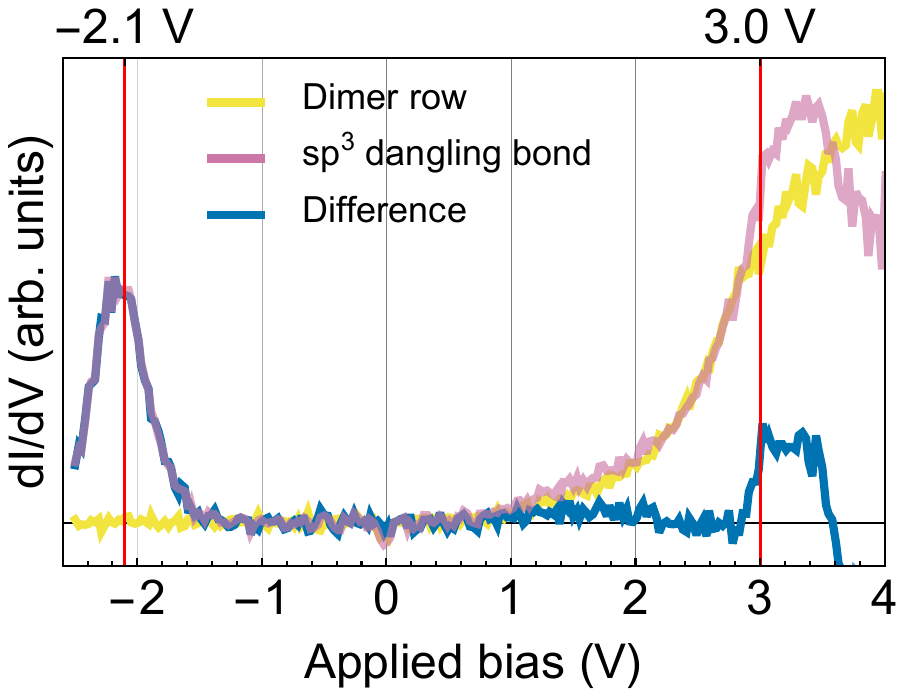}
        \caption{\justifying The $dI/dV$ spectrum taken directly above an sp${}^3$ dangling bond (pink) and the dimer rows of the pristine reconstructed H--C(100):$2\times1$ surface (yellow). Electronic states of the surface obscure the spectrum of the dangling bond. Consequently, the difference between the yellow and pink spectra provides a first-order approximation of the electronic features associated with the sp${}^3$ dangling bond (blue).}
    \end{subfigure}
\caption{\justifying Experimental STM and STS of the (100) diamond surface.}\label{fig:2}
\end{figure}

High-resolution STM imaging was first used to identify an sp${}^3$ dangling bond on the H--C(100):$2\times1$ surface. A constant-current topograph of the diamond surface after annealing at 823~K for an hour is presented in Figure~\ref{fig:2}~(a). Different defects can be observed on the surface and identified based on the symmetry of their STM topographies\cite{Sung2025}. The sp${}^3$ dangling bond is highlighted in blue and is characterised by a bright, chestnut-shaped feature. This identification was confirmed post STS by re-passivating the dangling bond through a targeted voltage sweep. Known as hydrogen `capping', this technique is an additional means to identify dangling bonds as documented by Sung~\textit{et al.}\cite{Sung2025}. 

The electronic structure of this particular sp$^3$ dangling bond was then studied using STS. It was chosen due to its isolation from other defects and step edges which could interfere with its electronic properties. For example, adjacent defects within 1--2 lattice sites can modify the orbital energy or wavefunction through direct electronic coupling or geometric distortion arising from electrostatic repulsion. Moreover, nearby defects could also produce an erroneous tunnelling current which obscures the spectrum of the target defect. While our simulation method could account for these effects through an appropriate first-principles model, here we focus explicitly on a spatially isolated sp${}^3$ defect.

A single-point $dI/dV$ spectrum was acquired on the sp${}^3$ dangling bond over a voltage range between $-2.5$~V and $+4$~V (sample bias). Attempts to extend the voltage range beyond these limits resulted in tip instabilities characterised by erratic (and non-reproducible) changes in tunnelling current, tip damage, or surface modification. We hypothesise that these effects are associated with large electric fields causing atomic rearrangement of the tip apex, or capping/desorption events on the surface. The spectrum is presented in Figure~\ref{fig:2}~(b) in pink. Two distinct peaks are present and centred at $-2.1$~V and $+3.4$~V. However, these features cannot be exclusively attributed to the sp${}^3$ dangling bond. This is because first-principles calculations of the H--C(100):$2\times1$ surface demonstrate that surface carbon also contributes to the electronic structure at the valence band maximum (VBM)\cite{Oberg2024}. Hence, the measured spectrum above the defect is expected to contain contributions from both the dangling bond and the underlying dimer-row electronic structure.

We separate these contributions by acquiring a reference STS spectrum. This is taken on a pristine dimer row near the dangling bond using the same tunnelling setpoint. The reference spectrum is presented in Figure~\ref{fig:2}~(b) in yellow. We then subtract the dimer-row spectrum from the dangling bond spectrum to produce the blue difference spectrum in Figure~\ref{fig:2}~(b). Following the tunnelling theory of Tersoff and Hamann, this difference may be interpreted as the first-order change in the local tunnelling density of states introduced by the dangling bond~\cite{TersoffHamann1985,Chen2021}. This neglects higher-order contributions arising from defect-induced perturbation of the surrounding surface states. However, these are expected to be small; the dangling bond is strongly localised and spatially isolated and therefore the extended electronic structure of the neighbouring dimer rows should be relatively unchanged. Consequently, the difference spectrum identifies the dominant spectral features associated with the dangling bond.

The difference spectrum shows two distinct features which can be attributed to the dangling bond at positive and negative applied bias. Preempting the findings from our first-principles studies below, these features correspond to tunnelling into distinct electronic states associated with the sp${}^3$ dangling bond. A key component of our analysis in Section~\ref{sec:disc} depends on identifying the applied bias which aligns the tip Fermi level with these electronic states. Tunnelling theory suggests that this should occur at the maxima of the features observed in the difference spectrum of Figure~\ref{fig:2}~(b). These occur at approximately $-2.1$~V and $+3.0$~V and are termed the resonance biases. The central aim of this work is to characterise the defect states associated with these resonance biases through first-principles calculations and electrostatic simulations.

\subsection{Electronic structure}

\begin{figure}[tbp]
   \centering
    \begin{subfigure}[b]{0.45\textwidth}
    \includegraphics[width=0.68\textwidth]{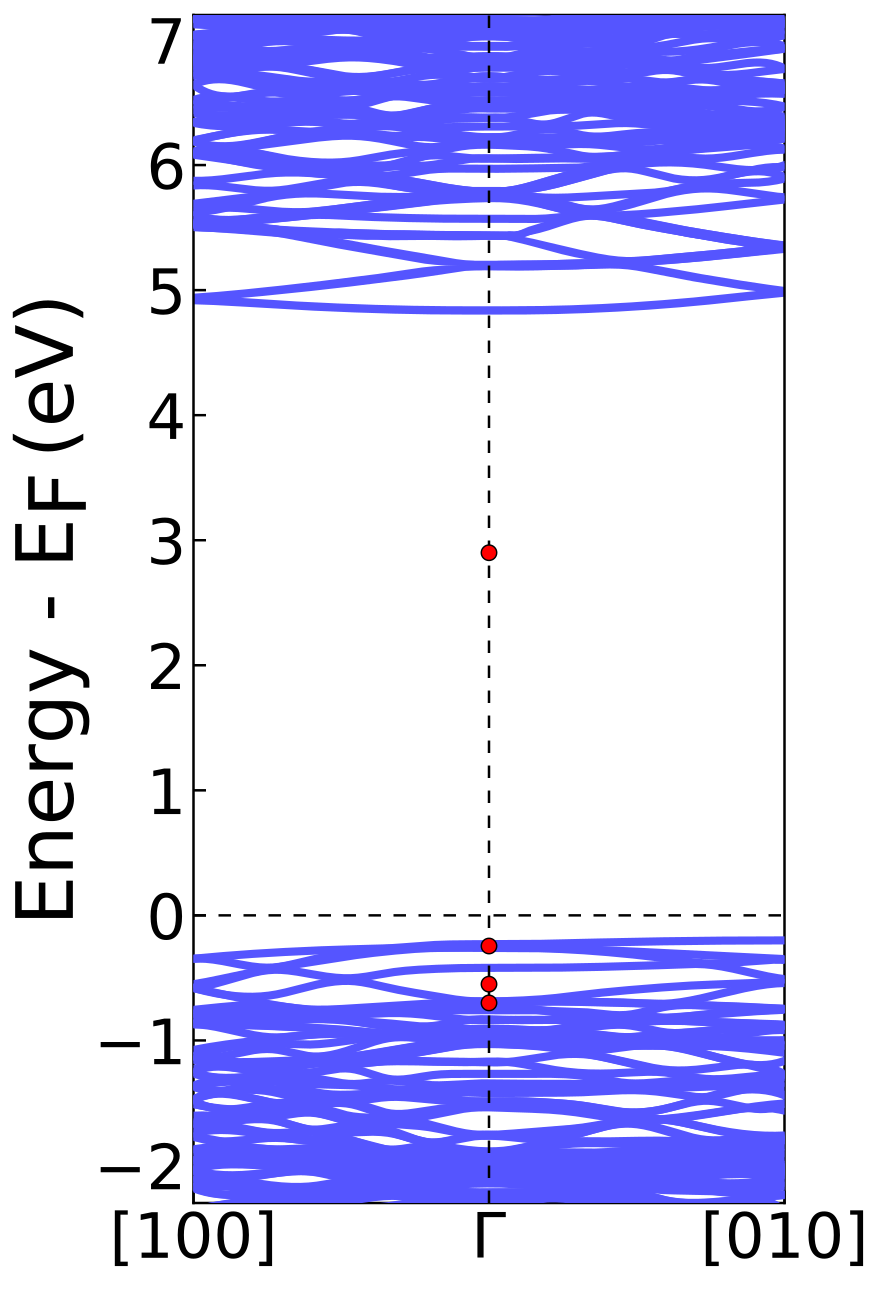}
        \caption{\justifying Band structure of the H--C(100):$2\times1$ surface with the energies of key sp${}^3$ defect states superimposed. The band structure corresponds to the diamond surface in the absence of any defect and is calculated using the PBE functional following the methodology in the supporting information of Ref.~\cite{Sung2025}. Note that the $\Delta$SCF method has been used to shift the valence and conduction bands to approximate the experimental band gap\cite{PhysRevB.86.035134}. The red dots are the $\Gamma$-point energies of the four key defect states of the sp${}^3$ dangling bond identified in the text. These were calculated using the HSE06 functional and the slab model in the absence of an electric field. These states have been aligned relative to the PBE band structure through occupied orbital 3, which resides at the surface VBM.}
    \end{subfigure}

    \begin{subfigure}[b]{0.45\textwidth} 
    \includegraphics[width=\textwidth]{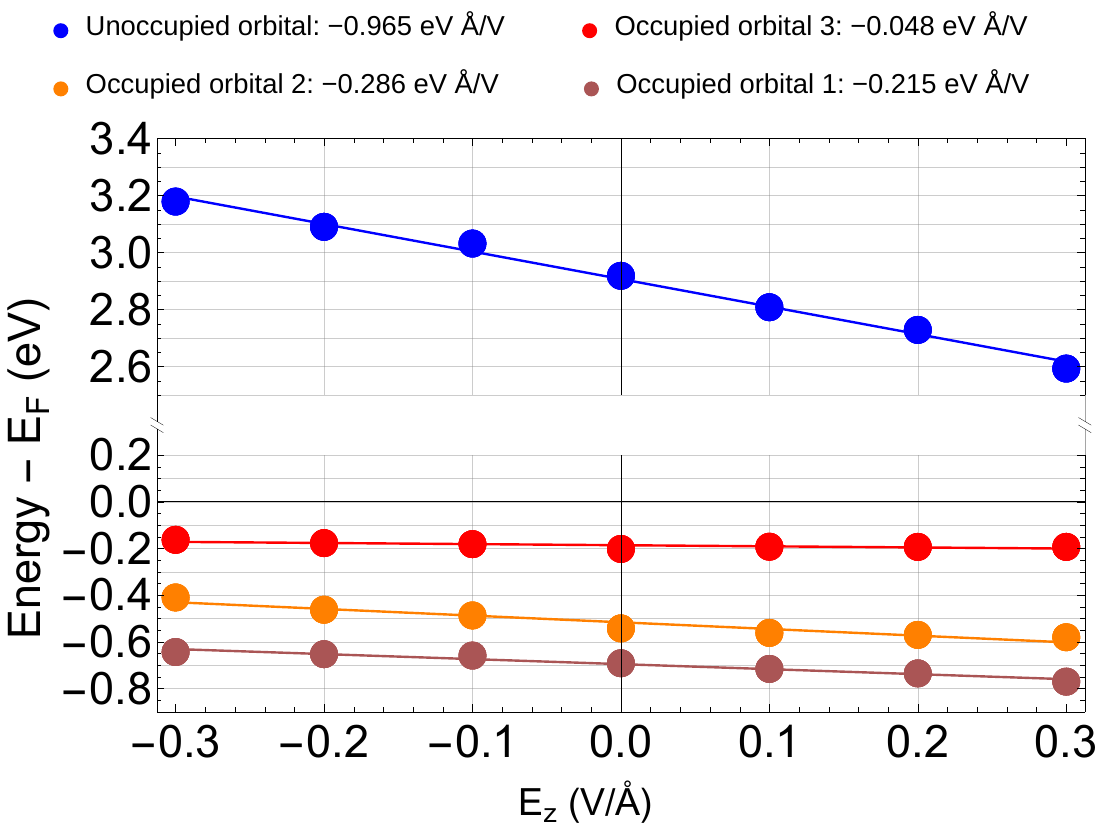}
        \caption{\justifying Stark shift of localised sp${}^3$ states at the $\Gamma$-point calculated using the HSE06 functional. This shows how the red dots in Figure~\ref{fig:3}~(a) change in energy under an applied electric field.}
    \end{subfigure}
    \caption{\justifying One-electron energies for the occupied and unoccupied orbitals for key states of the sp${}^3$ dangling bond on the H--C(100):$2\times1$ surface relevant to STM.}
    \label{fig:3}
\end{figure}

The electronic band structure of the H--C(100):$2\times1$ diamond slab (in the absence of the sp${}^3$ defect) is presented in Figure~\ref{fig:3}~(a). This has been generated using the PBE functional with band-gap energies corrected using the $\Delta$SCF method\cite{PhysRevB.86.035134}. Superimposed on the band structure are key HSE06 $\Gamma$-point energies for the H--C(100):$2\times1$ slab with the sp${}^3$ defect. These states of interest are those associated with the sp${}^3$ dangling bond near the bulk Fermi level ($E_F$). For boron-doped diamond, $E_F$ is positioned approximately $+0.2$~eV above the bulk VBM. The defect-related states have been identified through explicit visualisation of charge densities at the $\Gamma$-point obtained using the HSE06 functional (see Figure~\ref{fig:1}). This reveals four key states relevant to STS which are associated with the defect electronic structure. There are three occupied states positioned at approximately $-0.7$~eV, $-0.54$~eV, and $-0.2$~eV relative to $E_F$ and an unoccupied state located $+2.9$~eV above $E_F$. For reference, the occupied states will be labelled from 1 to 3 in ascending order of energy. It is assumed that the two peaks in the difference spectrum can be attributed to tunnelling involving at least two of these four states.

The diamond surface experiences a considerable electric field during STS. This produces a linear Stark shift in the energy of the defect states which must be accounted for when analysing the STS spectrum. This Stark shift is demonstrated explicitly in Figure~\ref{fig:3}~(b), which presents the energy of the four key defect states as a function of electric field. The charge density of the unoccupied state extends further into the vacuum region than the occupied states, and therefore experiences the greatest Stark shift.

\subsection{Band bending}

Application of a bias voltage during STM/STS translates the tip Fermi level relative to the bulk electronic band structure. However, unlike a metallic surface, the electric field produced by the tip is not completely screened at the diamond surface. Due to diamond's wide band gap there is no accumulation of intrinsic charge carriers at the surface. This can be contrasted to the behaviour of an ideal capacitor. In diamond, the applied bias between tip and sample is not fully dropped across the vacuum region. Consequently, for an applied bias $V_\text{bias}$, the tip Fermi level is \textit{not} translated by energy $qV_\text{bias}$ relative to the diamond surface states. Instead, it is translated by a reduced amount $qV_\text{drop}$, where $V_\text{drop} = V_\text{bias} - V_\text{bb}$ for $V_\text{bb}$ the band bending potential at the surface.

This means that the resonance biases in Figure~\ref{fig:2}~(b) cannot be directly compared to the electronic band structure in Figure~\ref{fig:3}~(a). One must also account for the band bending produced by the reduced screening response at the diamond surface. This screening is asymmetric depending on the polarity of the bias. At positive sample bias, holes accumulate at the surface and provide a strong screening response. At negative sample bias the screening response is due to ionised boron acceptors, which are deep defects with an activation energy of approximately $0.4$~eV. In contrast to shallow dopants in silicon, ionisation is incomplete in boron-doped diamond and the screening response is poor. Large negative biases are therefore required to translate the tip Fermi level relative to the defect states.

Fortunately, the voltage drop and surface electric field can be estimated using electrostatic modelling. In our circumstance, this modelling depends on three key variables within Poisson's equation and its boundary conditions: the acceptor concentration ($N_A$), the absolute tip-surface separation ($t_H$), and the built-in voltage ($V_\text{BI}$). The latter is defined by the difference in work functions between tip and sample, and can be calculated experimentally through measuring the function $I(t_H)$. This is performed in the Supporting Information and yields a value of $-0.9$~eV. As discussed in Section~\ref{sec:methods}, for our sample we estimate that  $10^{16} \ \text{cm}^{-3} \leq N_A \leq 10^{20} \ \text{cm}^{-3}$. Finally, $t_H$ is generally unknown during STM/STS and difficult to quantify, but typically lies within the tunnelling regime of $5$ -- $10$~\AA. Both $N_A$ and $t_H$ are therefore unknown (but constrained) variables, and the goal of Section~\ref{sec:disc} is to determine them through fitting to the STS spectrum in Figure~\ref{fig:2}~(b).

\begin{figure}[tbp]
    \centering
    \begin{subfigure}[b]{0.47\textwidth}
        \includegraphics[width=\textwidth]{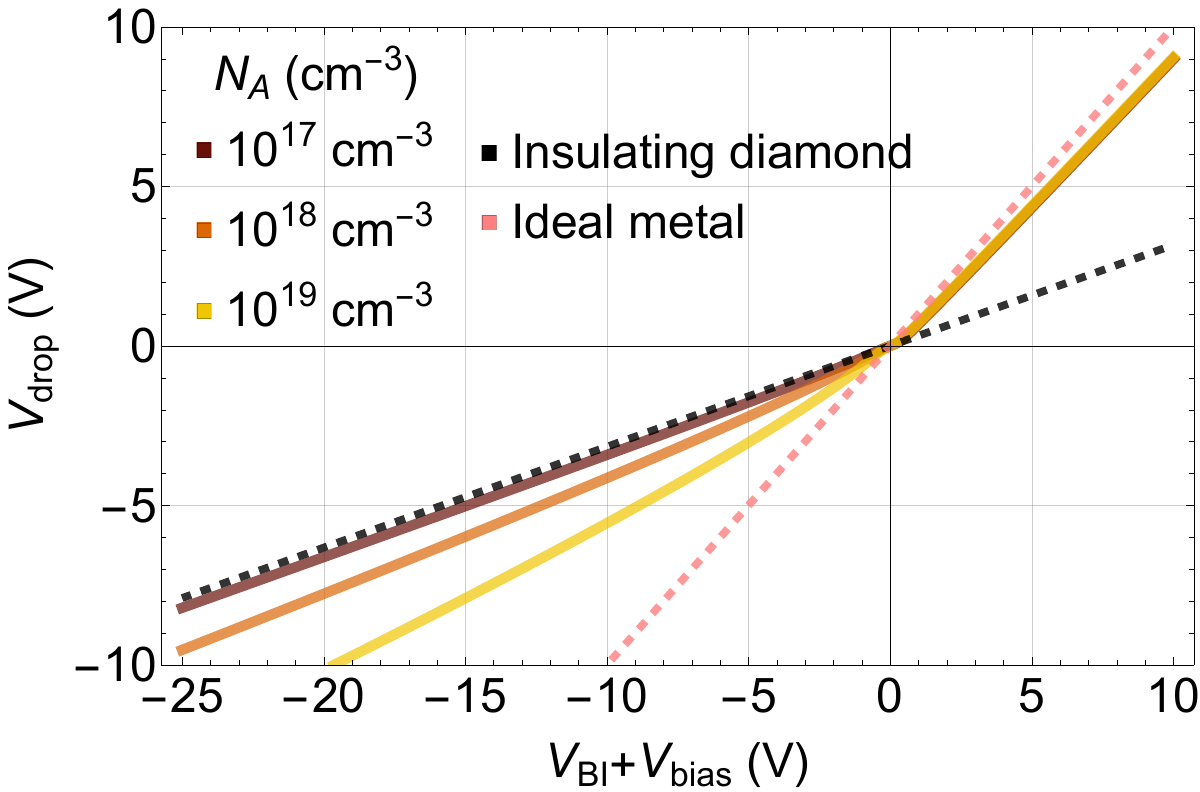}
        \caption{\justifying Voltage drop at the diamond surface.}
    \end{subfigure}
 \hspace{0.1\linewidth} 
    \begin{subfigure}[b]{0.47\textwidth}
        \includegraphics[width=\textwidth]{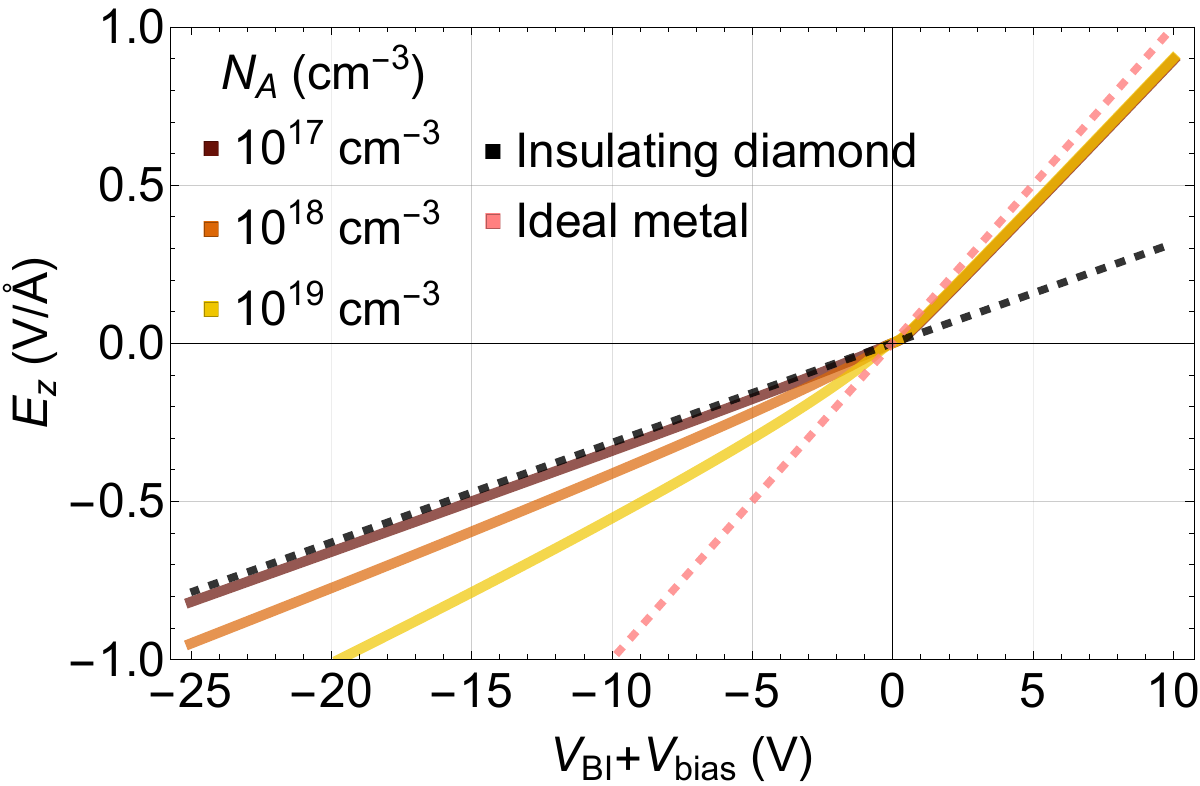}
        \caption{\justifying Electric field at the diamond surface.}
    \end{subfigure}
\caption{\justifying The voltage drop and electric field as a function of applied bias. Due to screening effects produced by holes and ionised acceptors, the tip Fermi level does not translate commensurately with the applied bias relative to the band structure of the sample surface. }\label{fig:4}
\end{figure}

The effects of band bending are demonstrated explicitly in Figure~\ref{fig:4}. We plot the relationship between $V_\text{drop}$ and the electric field at the surface as a function of $V_\text{bias}$ and $N_A$. The accumulation of holes at positive sample bias produces a screening response which is comparable to an ideal metal surface ($V_\text{bias}=V_\text{drop}$). However, at negative sample bias the magnitude of the vacuum voltage drop is much smaller than that of the applied bias. Peaks at negative bias in the $dI/dV$ spectrum therefore correspond to states which are \textit{higher} in energy within the surface electronic structure. Note that the screening response decreases as $N_A$ decreases due to a lower density of ionised defects, and increases with $t_H$ due to a greater voltage drop occurring across an extended vacuum region.

\section{Discussion}
\label{sec:disc}

The defect states contributing to the difference spectrum in Figure~\ref{fig:2}~(b) can now be elucidated. Firstly, the Stark shifts in Figure~\ref{fig:3}~(b) describe how the defect state energies are modified due to an applied electric field. Secondly, the electrostatic modelling presented in Figure~\ref{fig:4} describes the electric field produced by an applied bias, and how this applied bias translates the tip Fermi level relative to the defect states. Through fitting the unknown variables $N_A$ and $t_H$, these two results uniquely determine which states produce the experimental peaks at $-2.1$~V and $3.0$~V.

\begin{figure}[tbp]
    \centering
    \begin{subfigure}[b]{0.47\textwidth}
        \includegraphics[width=\textwidth]{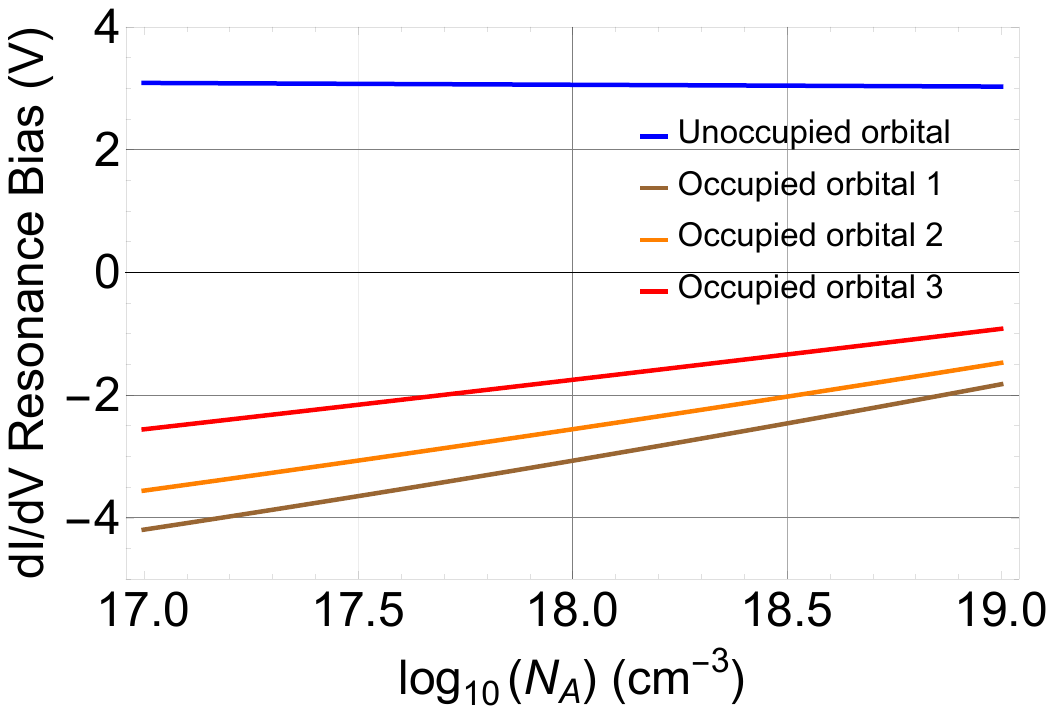}
        \caption{\justifying Position of resonance biases in the difference spectrum as a function of $N_A$ ($V_\text{BI}=-0.9$~V, $t_H=7.5$~\AA).}\label{fig:pNa}
    \end{subfigure}

    \begin{subfigure}[b]{0.47\textwidth}
        \includegraphics[width=\textwidth]{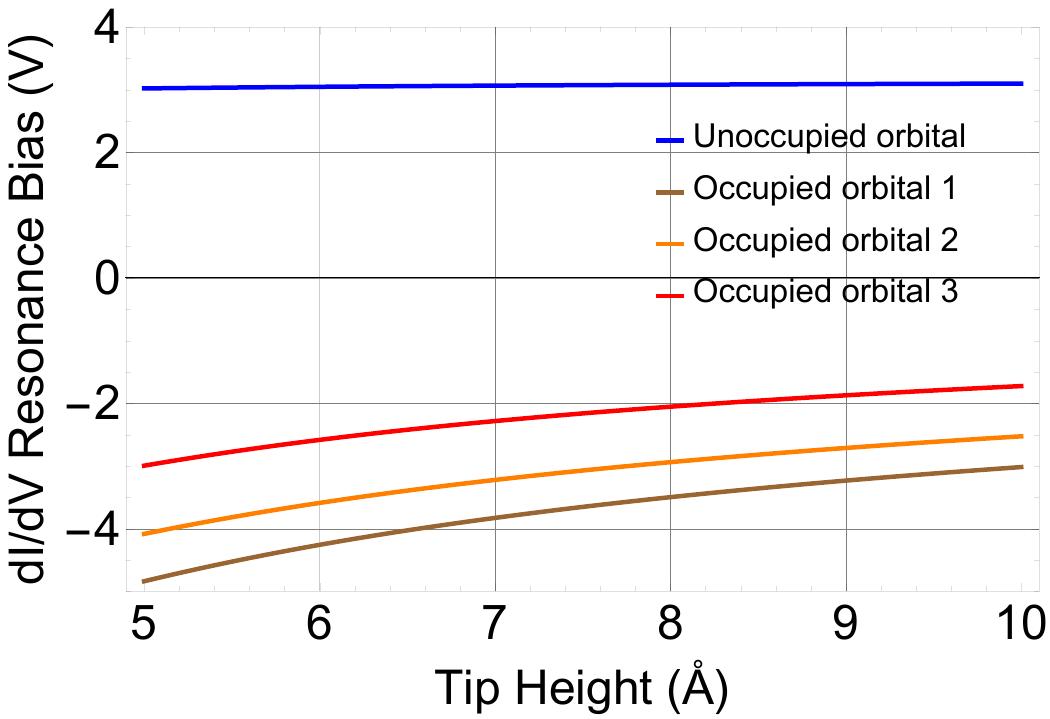}
        \caption{\justifying Position of resonance biases in the difference spectrum as a function of $t_H$ ($V_\text{BI}=-0.9$~V, $N_A=10^{17.5}$~cm${}^{-3}$).}\label{fig:pHt}
    \end{subfigure}

    \begin{subfigure}[b]{0.47\textwidth}
        \includegraphics[width=\textwidth]{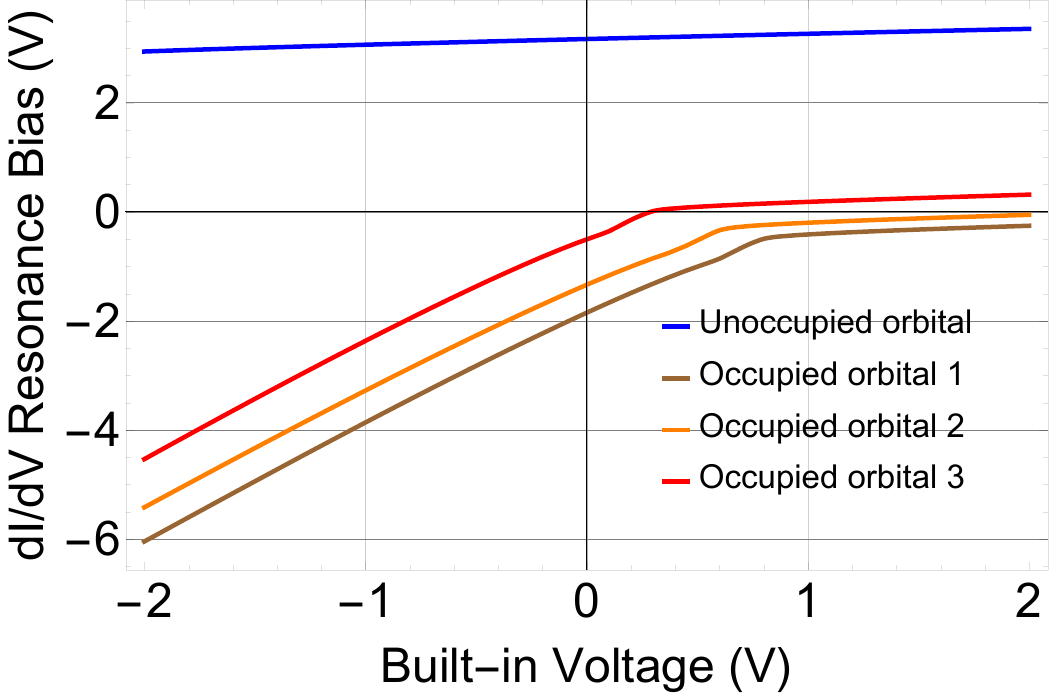}
        \caption{\justifying Position of resonance biases in the difference spectrum as a function of $V_\text{BI}$ ($t_H=7.5$~\AA, $N_A=10^{17.5}$~cm${}^{-3}$).}\label{fig:pBI}
    \end{subfigure}
\caption{\justifying Theoretical peak positions for each of the four electronic states in the difference spectrum under different electrostatic conditions.}\label{fig:5}
\end{figure}

To make this fitting procedure explicit, consider the variation of features in the difference spectrum as a function of the model parameters. Figure~\ref{fig:5} depicts the theoretical resonance biases for the four localised states as a function of $N_A$, $t_H$, and $V_\text{BI}$. The effect of each variable has been considered in isolation by fixing the other two variables to realistic values. Two key results can be inferred from Figure~\ref{fig:5}. Firstly, the bias required to tunnel into the unoccupied orbital is effectively fixed at 3~V and varies only weakly with $N_A$, $t_H$, and $V_\text{BI}$. As expected, band bending is minimal due to the effective screening of the mobile holes at positive bias. In contrast, the occupied orbitals experience significant band bending due to the ineffective screening response of the ionised acceptors. Secondly, under many conditions the three occupied orbitals are close together, particularly for $V_{BI} > 0$. However, the experimental spectrum in Figure~\ref{fig:2}~(b) displays a single peak at negative voltage with FWHM of approximately 0.5~V. At the representative settings $N_A=10^{17.5}$~cm${}^{-3}$, $t_H=7.5$~\AA, and $V_\text{BI}=-0.9$~V in Figure~\ref{fig:5}, occupied orbitals 1 and 2 lie below the measured voltage range. This supports the assignment in which the two peaks in Figure~\ref{fig:2} can be associated with tunnelling into the unoccupied orbital and out of occupied orbital 3.

\begin{figure}[tbp]
    \centering
\includegraphics[width=0.95\linewidth]{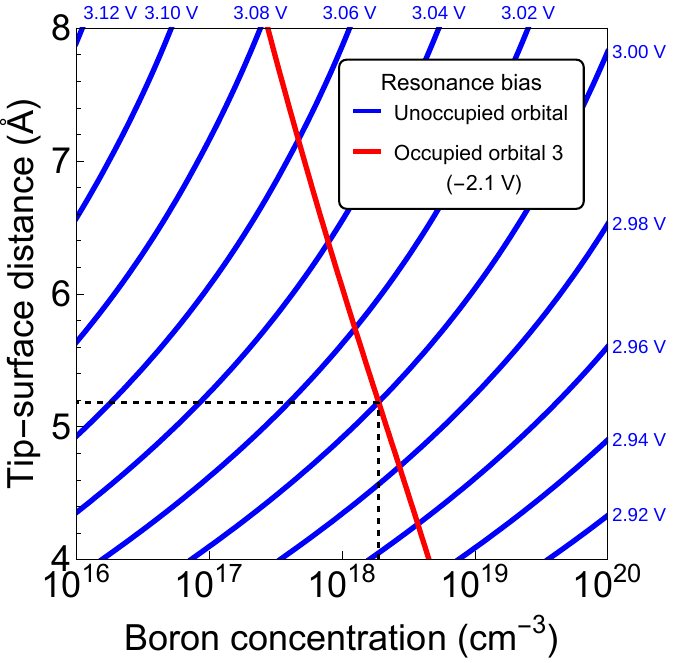}
\caption{\justifying Contour plot for resonance biases corresponding to the unoccupied orbital and occupied orbital 3 at $V_\text{BI}=-0.9$~V. The resonance at negative bias is strongly dependent on both $t_H$ and $N_A$, whereas the resonance at positive bias is only weakly dependent. Consequently, uncertainty in the positive resonance bias produces a correlated range of tip--surface distances and acceptor concentrations. Assigning the resonance biases to the experimental values found in Figure~\ref{fig:2}~(b) of $-2.1$~V and $3.0$~V gives $t_H = 5.2$~$\text{\AA}$ \ and $N_A = 1.9 \times 10^{18} \ \text{cm}^{-3}$. However, an uncertainty of $\pm 0.05$~V in the positive resonance bias corresponds approximately to $t_H$ between $4.1~\text{\AA}$ and $6.8~\text{\AA}$.}
    \label{fig:6}
\end{figure}

The contour plot in Figure~\ref{fig:6} presents the possible values of $t_H$ and $N_A$ given resonance biases for the two orbitals. The red line represents the voltage for occupied orbital 3 which is fixed at $-2.1$~V. The blue lines represent a range of possible voltages for the unoccupied orbital between $2.92$~V and $3.12$~V. The positions where the red line intersects the blue lines represent valid values for $t_H$ and $N_A$. If we choose the resonance bias for the unoccupied orbital as $3.0\ \text{V}$, this yields $t_H = 5.2\ \text{\AA}$ \ and $N_A = 1.9 \times 10^{18}\ \text{cm}^{-3}$. However, a small variation of $\pm 0.05\ \text{V}$ in the unoccupied peak position corresponds approximately to $t_H$ between $4.1\ \text{Å}$ and $6.8\ \text{Å}$ in the displayed contour family. Ambiguity in the resonance bias therefore translates into uncertainty in the absolute tip height.

This uncertainty does not represent a failure of our methodology. Instead, it reflects the broad width of the peaks in the difference spectrum and is further exacerbated by the asymmetry in band bending at positive and negative biases (namely, the energy of the unoccupied orbital is not strongly affected by changes in $t_H$ or $N_A$). Consulting Figure~\ref{fig:5}, more accurate values of $t_H$ and $N_A$ could be obtained by fitting to two occupied orbitals. This is because the resonance bias for both peaks in the difference spectrum would be strongly affected by band bending in boron-doped diamond. Unfortunately, we cannot test this hypothesis as we only observe a single feature at negative bias, and were unable to reliably extend the range of our voltage window below $-2.5$~V. Increasing the accuracy of our technique is therefore left for future work on diamond or other semiconducting surfaces.

\section{Conclusion}
\label{sec:conc}

Peak positions in STS spectra do not provide a reliable indicator of orbital energies for wide band-gap semiconductors. For boron-doped diamond, band bending decreases the energy of occupied surface states relative to the bulk Fermi level. Electrostatic modelling demonstrates that the extent of this band bending is dependent on three factors: the doping concentration, the tip height, and the built-in voltage. Fortunately, first-principles techniques can calculate the variation in defect energies due to these factors. Given an STS spectrum of a given defect, it is then possible to fit these factors by fitting.

This procedure has been implemented for the sp${}^3$ dangling bond on H--C(100):$2\times1$ to obtain a comprehensive understanding of its STS signature. We identify two key features in the $dI/dV$ curve. The first is a resonance at $+3.0$~V corresponding to the mid-gap unoccupied orbital that is resistant to band bending. We note that previous work indicates that many other diamond defects also possess unoccupied mid-gap states near 3~eV\cite{Sung2025}. The second is a resonance at negative bias corresponding to an occupied orbital near the surface VBM; its position varies with dopant concentration, tip height, and built-in voltage. The presence of these two states can aid in the identification of the sp${}^3$ dangling bond when combined with STM topography.

\section{Funding Information}
This research was funded by the Australian Government through an Australian Research Council Linkage Project (LP200301428). This work was performed in part at the Melbourne Centre for Nanofabrication (MCN) in the Victorian Node of the Australian National Fabrication Facility (ANFF). This research was undertaken with the assistance of resources and services from the National Computational Infrastructure (NCI), which is supported by the Australian Government.

\clearpage

\renewcommand{\bibsection}{\section*{References}}

\bibliographystyle{unsrt}
\bibliography{biblio}

\clearpage
\onecolumngrid
\setcounter{figure}{0}
\setcounter{equation}{0}
\renewcommand{\thefigure}{S\arabic{figure}}
\renewcommand{\theequation}{S\arabic{equation}}
\providecommand{\theHfigure}{}
\providecommand{\theHequation}{}
\renewcommand{\theHfigure}{supp.\arabic{figure}}
\renewcommand{\theHequation}{supp.\arabic{equation}}
\begin{center}
{\Large\bfseries Supporting Information}\par\medskip
Tunnelling probe-based characterisation of the sp${}^3$ dangling bond\\
on the H--C(100):$2\times1$ surface
\end{center}
\section*{Built-in voltage determination}

The built-in voltage difference ($V_\text{BI}$) may be calculated as the difference between the work function of the sample ($\Phi_s$) and tip ($\Phi_t$) surfaces. The work function of the H--C(100):$2\times1$ surface can be determined from its electron affinity ($\chi=-1.3$~eV), bulk band gap ($E_g = 5.5$~eV), and Fermi level relative to the bulk valence-band maximum ($E_F \approx 0.2$~eV for boron-doped diamond). This yields $\Phi_s = E_g + \chi - E_F \approx 4.0$~eV. 

The tip work function cannot be determined in such a straightforward manner as its atomic composition and hence its electronic structure are unknown. However, we estimate the tip work function from $I(\Delta t_H)$ using the following model. The assumptions required are similar to those used in Chen's approximation for tunnelling matrix elements\cite{Chen1990,Chen2021}. Namely, the crystal potential from the sample and tip subsystems vanishes within the vacuum region. In this region, the Schr\"odinger equation for states of the tip apex ($\phi_\mu$) which are close in energy to the Fermi level reduces to the Helmholtz equation. That is,
\begin{equation}\label{helmholtz}
(\nabla^2 - \kappa^2)\phi_\mu(\mathbf{r}) = 0,
\end{equation}
where $\kappa=(2m\Phi_t)^{1/2}/\hbar$ and $m$ is the electron mass. The solutions to equation~\eqref{helmholtz} decrease exponentially as a function of radial distance from the tip apex. Calculation of the tunnelling matrix elements using Bardeen's theory yields the following dependence of current with changing tip height,
$$I(\Delta t_H) \propto \exp(-2\kappa \cdot \Delta t_H).$$
Note that $\Delta t_H$ is not the absolute tip height itself (which is not necessarily known), but instead represents the change in tip height above the surface within the tunnelling regime. Hence, experimental measurement of $I(\Delta t_H)$ can be used to determine $\kappa$ and therefore $\Phi_t$.

\begin{figure}[ht]
    \centering
\includegraphics[width=0.65\linewidth]{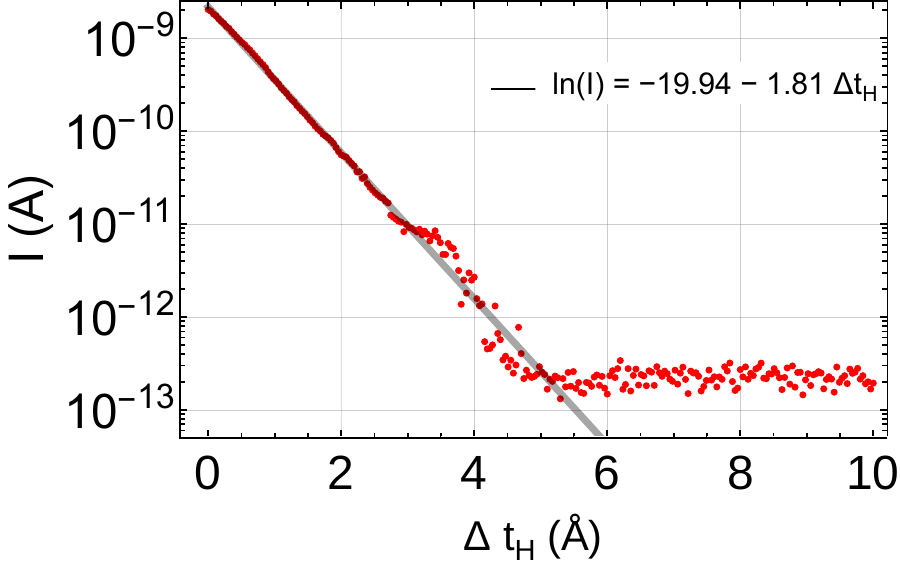}
\caption{\justifying Logarithmic plot of current versus change in tip-surface separation distance recorded on the H--C(100):$2\times1$ surface at 1.5~V using a Ag/W tip. A linear fit over 0--5~\AA\ gives $\kappa=0.9$~\AA${}^{-1}$ and $\Phi_t=3.1$~eV.}
    \label{fig:SI}
\end{figure}

Figure~\ref{fig:SI} presents a logarithmic plot of $I(\Delta t_H)$ sampled above the H--C(100):$2\times1$ dimer rows at an applied bias of 1.5~V. The tunnelling regime can be identified from the exponential decay between $0~\text{\r{A}} <\Delta t_H \lesssim 5~\text{\r{A}}$. Note that these values do not correspond to an absolute tip height, but simply represent the relative change in tip height throughout the experiment. A linear fit gives $\kappa=0.9$~\AA${}^{-1}$, which corresponds to $\Phi_t = 3.1$~eV and $V_\text{BI} \approx -0.9$~eV. Consequently, at zero applied bias the Fermi level of the two subsystems will be equalised through the migration of holes from sample to tip. The diamond surface states will bend downwards, requiring a lower applied bias to achieve tunnelling out of the occupied states.

\clearpage

\end{document}